\documentclass[vecphys]{svmult}

\usepackage{graphicx}        
\usepackage[bottom]{footmisc}
\def\ts{\times}
\def\beq{\begin{equation}}
\def\ee{\end{equation}}
\def\ep{\epsilon}
\def\apjl{ApJL}
\def\apj{ApJ}
\def\aap{A \& Ap}
\def\mnras{MNRAS}

\begin{document}

\title*{Planetary  Nebulae  Principles \& Paradigms: 
Binaries, Accretion, Magnetic Fields}
\titlerunning{Pre-Planetary/Planetary Nebulae Engine Principles and Paradigms}
\author{Eric G. Blackman \inst{}\and
Jason T. Nordhaus\inst{}}
\institute{Department of Physics and Astronomy, University of Rochester,
Rochester, NY, 14627, USA, 
\texttt{blackman@pas.rochester.edu}}
%
%
\maketitle

\begin{abstract}
Observations suggest that many, if not all, post AGB systems evolve through 
an aspherical outflow phase. Such outflows require a sufficient engine 
rotational energy which binaries can provide. 
Via common envelope evolution, binaries can 
directly eject  equatorial outflows or produce
poloidal outflows from magnetized accretion disks around the primary
or secondary.  We discuss how accretion driven magnetohydrodynamic
outflow models all make similar predictions for the outflow power and speed and 
we distinguish between the launch vs. propagation regimes
of such outflows. We suggest that 
the high velocity bipolar outflows observed in planetary nebulae (PNe)
and the lower velocity but higher power bipolar outflows observed in pre-PNe (pPNe) 
are  kinematically consistent with time dependent 
accretion onto a white dwarf (WD) within a depleting envelope.
Since the WD primary core is always present in all post-AGB systems, 
accretion onto this core is potentially common.
Previous work has focused on core accretion from sub-stellar companions,  
but low mass stellar companions may be more 
important, and further work is needed.

\end{abstract}

\section{Introduction:  Kinematics of pPNe and PNe}
\label{sec:1}

Understanding the origin of asymmetric outflows in PNe and pPNe requires
feedback between observers, specific object modelers, 
and paradigm-seeking, order of magnitude theorists. 
Here we behave as the latter.

Generally,  pPNe  exhibit
 a combination of a fast bipolar outflow embedded within a 
slow spherically symmetric wind from the AGB star \cite{bujar01}.
Presently, the data do not  rule out  all pPNe having gone through
a strongly asymmetric outflow and all PNe having gone through an 
asymmetric pNE phase. The symmetry of  PNe would then correlate with age
and the evolution from the pPNe to 
PNe could reflect a time evolution of the same physical mechanism 
that produces asymmetry on small scales but leads to  a
a nearly spherical structure on large scale as supersonic motions damp. 
While AGB stars produce spherically symmetric outflows, pPNe asymmetry
arises within $\le 100$ yr \cite{bujar01,vlemmings06}.

For pPNe \cite{bujar01}, each fast  wind has a typical age 
$\Delta t\sim 10^2-10^3$yr,  speed $\sim 50$km/s,
 mass $M_f\sim 0.5M_\odot$, outflow rate, ${\dot M}_f\sim 5 \times 10^{-4}M_{\odot}/{\rm yr}$, 
 momentum $\Pi\sim 5 \ts 10^{39}{\rm g.cm/s}$, and mechanical 
luminosity $L_{m,f}\ge 8\ts 10^{35}{\rm erg/s}$ (can be as high as 
$10^{37}$erg/s).
The slow pPNe wind has an age  $\Delta t\sim 6\ts 10^3$yr, a speed $v_w\sim 20$km/s 
a mass   $M_s\sim 0.5M_\odot$, outflow rate, ${\dot M}_s\sim 10^{-4}M_{\odot}/{\rm yr}$, 
 momentum $\Pi_s \sim 2 \ts 10^{39}{\rm g\ cm/s}$, and mechanical 
luminosity $L_{m,s}\sim  10^{34}{\rm erg/s}$.

For PNe observations suggest e.g. \cite{balickfrank02}
an  age 
$\Delta t\sim 10^4$yr 
a slow wind  of speed $v_s\sim 30$km/s 
of mass $M_s\sim 0.1M_\odot$, outflow rate, ${\dot M}_s\sim 10^{-5}M_{\odot}/{\rm yr}$, 
 momentum $\Pi_s \sim 6 \ts 10^{38}{\rm g\ cm/s}$, and mechanical 
luminosity $L_{m,s}\sim 3 \ts 10^{33}{\rm erg/s}$.
PNe have  fast  winds of speed $v_f\sim 2000$km/s,
 mass $M_f\sim 10^{-4}M_\odot$, outflow rate, ${\dot M}_f\sim 10^{-8}M_{\odot}/{\rm yr}$, 
 momentum $\Pi_f\sim 4 \ts 10^{37}{\rm g\ cm/s}$, and mechanical 
luminosity $L_{m,f}\sim 1.3 \ts 10^{34}{\rm erg/s}$.  

The pPNe phase  demands the  most power. 
The linear momenta of fast bipolar pPNe
outflows seems  too large for radiation driving \cite{bujar01}. This 
motivates the need to tap rotational energy 
which binaries can  provide. The 
 high fraction of binaries attributed to p/PNe 
has led to speculation that all asymmetric  PNe and pPNe involve
binaries \cite{moe06,soker06}.

How is the rotational energy and angular momentum
converted into outflows?  Differential
rotation supplied by binaries can amplify magnetic fields which can
in turn produce accretion powered bipolar jets.
Accretion powered, magnetically mediated jets, are seemingly ubiquitous
in astrophysics and can 
accommodate the high momentum pPNe demands \cite{bfw01}.






%
%
%


\section{Which Binary Accretion Scenario?}

In a common envelope (CE) \cite{ibenlivio} the companion drags on
the envelope of the primary, transferring angular momentum and kinetic energy. 
Provided that the envelope cooling time scale 
exceeds the energy supply time from the in-spiraling companion, 
a converted fraction $\alpha \ge 0.1$ of its loss in gravitational energy
can spin up the envelope and unbind it.
For a fixed envelope mass, a lower mass companion must fall deeper
 to unbind the envelope.
Fig. 1  \cite{nb06} shows aspects of 
CE  for low mass companions  in  an AGB envelope. 
When the Roche radius is reached by the in-spiraling companion,
accretion can occur.

\begin{figure}
\centering
\includegraphics[height=5cm]{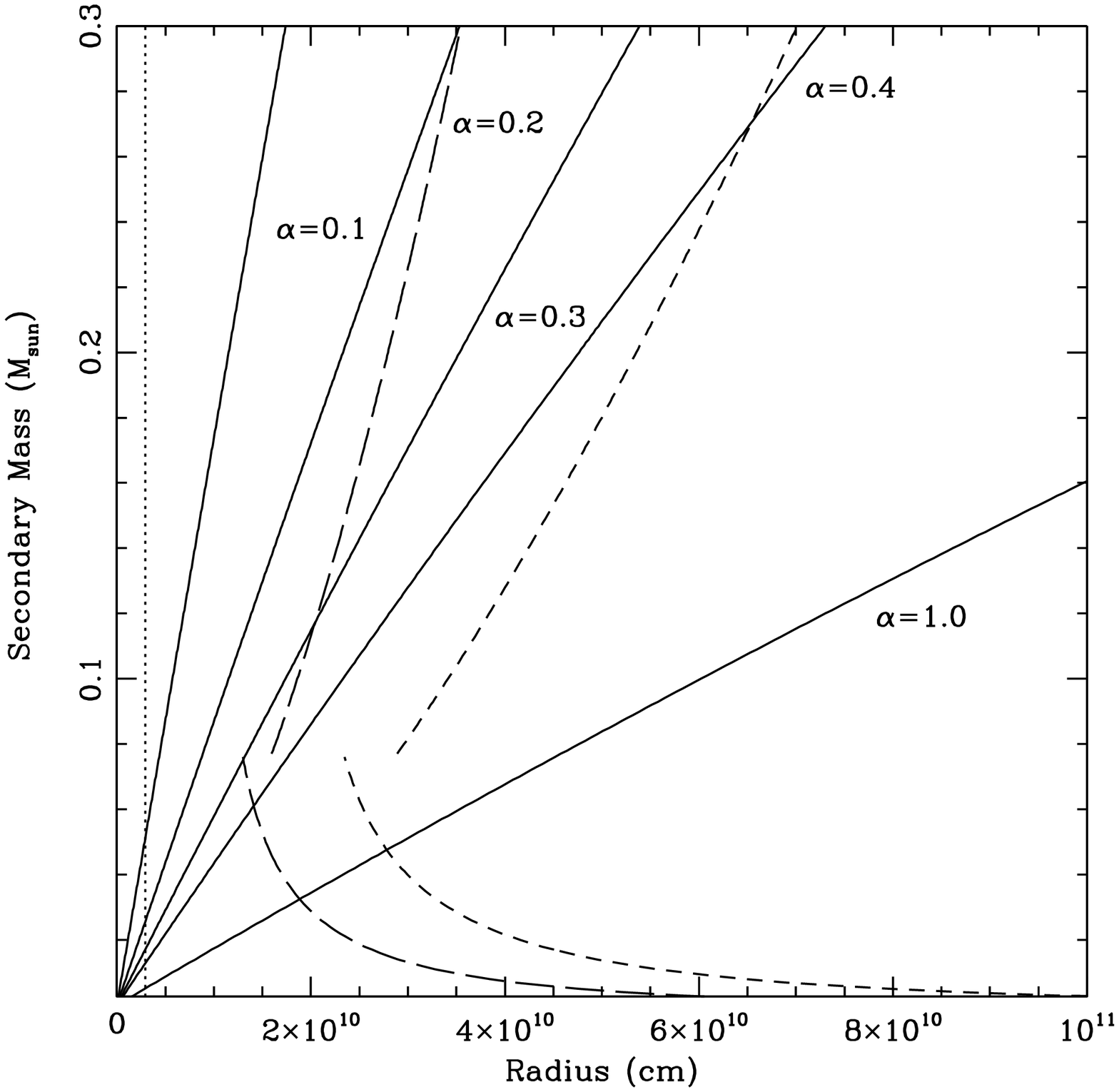}
\includegraphics[height=5cm]{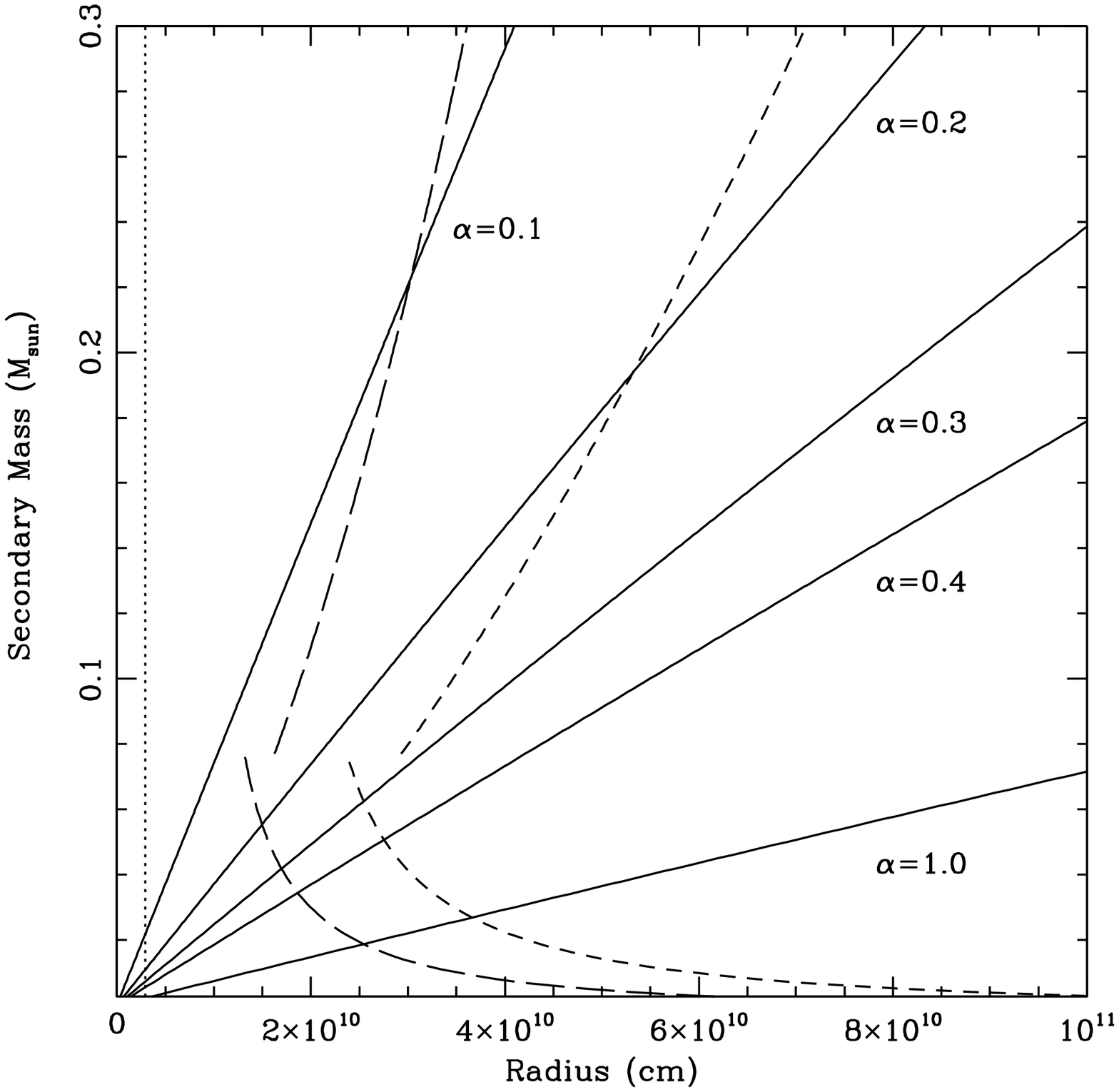}
\caption{From \cite{nb06}:
For a 3$M_\odot$ young AGB star (left) and inter-pulse AGB star (right), 
the solid lines show the radius where 
a companion must in-spiral 
 to unbind the envelope for different values of the drag parameter $\alpha$.  The dash-dotted line shows the core-envelope boundary. The long-dashed line marks the tidal shredding radius
and the short dashed line is where the companion first fills its Roche lobe.}
\label{fig:1}       
\end{figure}

Ref \cite{rrl99} discusses 
the   secondaries  
for which accretion disks will form around the primary core. 
Brown dwarf (BD) ($0.003M_\odot <M_{crit}\sim 0.07M_\odot$) 
radii increase with decreasing mass unlike their Roche
radii which decrease with decreasing mass.
Such objects will unstably lose mass. Since the
circularization radius lies outside of the primary's core, 
a disk can form within a few orbit times. 
This contrasts the $M>M_{crit}$ case for 
which the stellar radius
decreases with decreasing mass and more strongly so than the Roche
radius. Supercritical companions 
present a circularization radius {\it within} the
primary's core. Material leaving the secondary would 
then initially spiral swiftly into the primary rather than orbit quasi-stably.

Ref.\cite{rrl99} focuses on BDs ($0.003M_\odot< M_2< M_{crit}$), but accretion from low mass stars
 and planets warrant further study:
Fig.1 (left) shows that for low values of the (uncertain)
drag parameter $\alpha$, some low mass stars could tidally shred into a disk
before reaching the radius where they unbind the envelope and subsequent
in-spiraling is potentially slowed.
Fig. 1 also shows that planets will shred into a disk 
upon inspiral and for large $\alpha$ in the inter-pulse AGB,
even large planets might unbind the envelope. 
Generalizations must incorporate   
the fact that the companion size  can be  
of order the  tidal shredding or Roche lobe radius, and the 
structural evolution of  the primary as the secondary in-spirals.
The latter is important for radii both exterior and interior to the companion:
How fast do further inspiral and
angular momentum transfer subsequently occur once the outer envelope is ejected?

If a  stellar companion unbinds
the envelope at a radius with
 Roche lobe overflow but without tidal shredding,  
sustained accretion requires the binary to lose
angular momentum. 
As the  secondary fills its Roche lobe, it will 
drag on any  residual inner envelope material. This  could 
transfer the needed 
angular momentum,  but too much drag could prevent  the companion 
from forming a quasi-stable Keplerian disk.
Material that initially leaves the companion for the primary
also carries magnetic fields and thus will magnetically link the primary and secondary.  Even though the circularization radius for $M_2 > M_{crit}$ 
is inside the core, the in-spiraling material
still incurs differential rotation.
Magnetic fields can then be amplified  
and can also release angular momentum.

Much of the initial accreted energy 
in the $M_2 > M_{crit}$ case would be released upon material impact to 
stellar surface.  This could produce dwarf novae type bursts
shrouded by the stellar envelope.
However, if the right range  of  mass and angular momentum are  transferred 
in this initial accretion phase to (1) drop  $M_2$   below $M_{crit}$,
(2) keep $M_2$ filling its Roche lobe, and  (3)  leave enough angular
momentum to form a Keplerian disk, then  
accretion  could   proceed  as for the initial $M_2<M_{crit}$ 
case of Ref.\cite{rrl99}.  Ref. \cite{huggins} suggests a 
$\sim O(100)$ year delay in the time scale between 
ejection of  cicumbinary dust tori and presence of jets in pPNe.
If the companion initially has $M_2> M_{crit}$, 
the fraction of the envelope ejected as it spirals
in might become the dust torus, and the delay before the
jet could be the time  it takes for the companion to lose
enough mass to move the  circularization radius
outside the core.

The role of  low mass stellar and planetary  companions
is important because although
$\sim 16\%$ of nearby ($<50$pc) sun-like (late F to early K) 
stars have companions more massive than Jupiter  at $<3$AU, 11\% are stars, 
5\% are planets, and $0\%$ are BDs  \cite{gl06}.
If  this $\sim 16\%$ were crudely taken as the fraction of 
binary induced, and thus asymmetric p/PNe, 
this is within 
a factor of 2 of estimates of the total fraction of low mass 
stars incurring any PNe \cite{soker06}. 

It is also possible for accretion disks to form around the secondary
\cite{sl94,mm98,soker05}.
The accretion rate inferred from the
Bondi wind accretion formulae is
\beq
{{\dot M}\over {\dot M}_s}=\left(M_2\over M_1\right)^2{(v/v_s)^4\over 
[1+(v/v_s)^2]^{3/2}},
\ee
where $v$ is the orbital speed of the secondary and
$v_w$ is the slow wind speed from the primary.
In general, for $M_2< M_1$, reasonable parameters
provide an accretion rate compatible with the luminosities
required at the PNe stage if the companion is a main sequence
star. However, the outflow velocities of the fast wind in PNe
would require a WD companion, as seen in the next section.
Complementarily, the luminosities of the fast wind pPNe outflows
would also require a WD companion.


\section{Accretion Disk Outflows in pPNe and PNe:}
\label{sec:5}

Accretion disk outflows have a mechanical luminosity of order
 \cite{bfw01}
\beq
L_{m}\sim {GM_*{\dot M}_{a}\ep \over 2 R_i} = 4.5 \ts 10^{36} \ep_{-1}{M_*{\dot M}_{-4}\over R_{i,10}},
\label{1}
\ee
where $\ep$ 
is the dimensionless 
efficiency of conversion from accretion to outflow,
$R_i$  is the inner disk radius, ${\dot M}_a$  is the accretion rate,
$G$ is Newton's constant, and $M_*$ is the central stellar mass.
In the last term  $R_{i,10}\equiv R_i/10^{10}{\rm cm}$, 
$\ep_{-1}\equiv \ep/0.1$ and ${\dot M}_{a,-4}= {\dot M}_a/10^{-4}M_\odot$/yr.
For an MHD outflow, Eq. (\ref{1}) equals the Poynting flux at the launch surface.

For propagation into a region of negligible inertia, 
the asymptotic outflow speed is $\sim \Omega r_A$  \cite{pudritz04}
where $\Omega$ is the angular speed
of field anchor point  and $r_A$ is the
radius where the  poloidal outflow speed 
equals the Alfv\'en speed. This product is typically of order  
 1-3 times the escape speed of the inner most radius of the disk and is thus at least 
\beq
v_{out}\sim v_{esc} =1600 \left(M_*\over R_{*,10}\right)^{1/2}{\rm km/s}.
\label{2}
\ee

Time dependent  accretion outflows described with 
Eqs. (\ref{1}) and (\ref{2}) are consistent  with the high pPNe outflow mechanical luminosity 
and the fast PNe wind speed  
of Sec. 1 when $M_*/R_*$ corresponds to a  WD 
(for accretion onto primary $M_*=M_1$, or secondary $M_*=M_2$)
The outflow speed from (\ref{2}) 
does not depend explicitly  on ${\dot M}_a$ 
(only implicitly and weakly 
via $R_i$ which can exceed the stellar radius for a strong field) and  is thus largely time independent.
However, this speed depends on the inertia of material 
blocking the outflow.  Conservation of momentum gives 
\beq
v_{obs}={M_{f}v_{f}
\over f_\Omega M_{env} + M_{f}}
\sim 80{\rm km/s},
\label{3}
\ee
where 
$M_{f}/ M_\odot=3.3 \times 10^{-4}\ep_{-1}{\dot M}_{a0,-3}\int_1^{1000} \tau^{-5/4}  d\tau
$ is the mass in one of the fast collimated  outflows, $M_{env}$ is the envelope mass,  $f_\Omega\sim 0.2$ is the solid
angle fraction intercepted by the collimated outflow, and
$\tau\equiv  t/1yr$ is used to incorporate
 ${\dot M}_a\propto t^{-5/4}$ of Ref. \cite{rrl99}.
The numbers have been  scaled to the pPNe case
 so that here $M_{env}>> M_{f}$ and for an envelope of mass $2 M_\odot$,
giving the intercepted mass of 0.2 $M_\odot$ for $f_\Omega=0.2$.
Eq. (\ref{3}) then
implies $v_{obs}=40$km/s, within the range observed \cite{balickfrank02}.

Eq. (\ref{3}) represents the observed speed of the fast 
when blocked and loaded by the 
envelope. By the end of the pPNe phase, the envelope is quite extended,  
reducing the optical depth and revealing material
moving at the ``free streaming'' fast wind speed. 
Assuming a dust-to-gass mass ratio of 1/100 and 
micron sized grains of density of 2g/cm$^3$, 
the optical depth from dust is 
\beq
\tau_d \sim 2.5 \ts 10^{-3}\left({n_d\over 2.5 \ts 10^{-13}{\rm cm^{-3}}}\right)
\left({\sigma_d\over  10^{-8}{\rm cm^2}}\right)
\left({R\over  10^{18}{\rm cm}}\right),
\ee
scaled for PNe.
For pPNe, the density increases by a factor $\ge 10^4$
and $R$ is down by a factor of 10, so $\tau_d\ge 2.5$,  optically  thick.
The different optical depths of pPNe and PNe  can thus explain
 why observed PNe fast winds can have 
$v_f>1600$km/s, whilst those of pPNe have $v_f<100$km/s. 

Keeping in mind our discussion of 
low mass stellar companions in Sec 2, note that Ref. \cite{rrl99}
considers  a  companion of mass $M_2\sim 0.03M_{\odot}<M_{crit}$ 
and a Shakura-Sunyaev viscosity parameter $\alpha_{ss}\sim 0.01$, 
for which the accretion rate then decays as 
${\dot M}_a\sim 1.6\ts 10^{-3}  t^{-5/4}{M_\odot}/$yr..
Using this in (\ref{1}) with $\ep=0.1$ 
for $t=100$ yr with $R_i= 2\times 10^9$cm
and $M_1=0.6M_\odot$ gives $L_{m,f}\sim 4.3 \times 10^{39}(t/{\rm 1yr})^{-5/4}$.
This provides the needed power demands of Sec. 1  for pPNe  after $1000$ yr and
for PNe after $10^4$ yr.  

Because the surface density evolves, the gas opacity
evolves from Thomson to Kramer's after $\sim 100$ yr \cite{rrl99}, 
and the  height to radius ratio decreases substantially as the 
disk cools. A fixed $\alpha_{ss}\sim 0.01$
is  self-consistent with the time evolving accretion and power
above.



\label{sec:5}

\section{Launch versus Propagation Regions for MHD Outflows}
\label{sec:1}

We refer to the ``launch'' region \cite{blackman07} 
of MHD  outflows
as the region where the magnetic force and energy dominates the flow 
and thermal energy. 
This extends to a height typically no greater than  
$z_c\sim 10-50 R_{i}$, where $R_{i}$ is the innermost radial scale of the engine.
(e.g. the inner radius of an accretion disk).
In the launch region the bulk flow is
accelerating but is sub-Alf\'enic until reaching $z_c$.
The ``propagation'' region describes $z>z_c$
where the poloidal flow speed exceeds the Alf\'en speed, 
eventually approaching its asymptotic speed.
Presently, only the propagation region is observationally  
spatially resolved.
With this distinction, we describe 3 classes of  MHD outflow related work.

{\it Launch to Propagation:} e.g. \cite{matt06,garcia05,anderson};  
Here the magnetic field is imposed 
to be ordered on a scale at least as
large  as the anchoring rotator. The base of the rotator
is typically a boundary condition.  Calculations address 
how material is accelerated by the combination of
centrifugal flinging of material along quasi-rigid field lines and/or 
a vertical gradient in the magnetic pressure. 
Both a poloidal and toroidal field component at the base are required.
The flow can be collimated and 
supersonic by hoop stresses before upon reaching the propagation region 
provided that there is an ambient pressure to collimate 
the magnetic field. Simulations can
cover from the base of the launch region to a scale typically $\le 100$
times the engine scale.

{\it Asymptotic Propagation:} e.g. \cite{garcia97,dennis07}; 
 Here a collimated jet is injected on small scales, and 
the subsequent propagation and shaping 
by the ambient medium of specified magnetic and thermal 
properties is studied. The  jet and ambient medium parameters 
are varied 
to assess what conditions can produce the observed asymptotic, morphologies.
These simulations do not address the magnetic field origin or 
acceleration mechanism.

{\it Field Origin to Launch:} e.g.\cite{bfw01,nbf07};
The   previous two categories do not
address where the dynamically important large scale
fields  comes from in the first place. Accretion of flux may be difficult in
a turbulent disk, but  these large scale fields 
can plausibly be produced by a combination of a flow dominated 
helical dynamo inside of the rotator, followed by magnetic buoyancy, and
a  magnetically dominated helical dynamo relaxation
in the corona \cite{blackman07b}. The latter opens up  
structures to the large scales needed to drive jets, much as solar coronal
loops open to create solar coronal holes. 
This category of work focuses on the field origin, with kinematic estimates of the subsequent launch, but not the dynamical
launch itself.

It is a  frontier  to couple the above
categories, and each   has  limitations when  separated from
 the other two.
For example,
a self-consistently
grown strong field dynamically mediating  the launch  need not  necessarily 
imply a strong magnetic influence in the propagation region:
If the launch region produces a super-magnetosonic collimated outflow, 
then even subsequent ballistic propagation into the propagation region
would still emerge as collimated.  In addition,  
a super-magnetosonic outflow could become turbulent and the
 turbulence can amplify small scale magnetic energy 
 to $> 10\%$ of  equipartition with this turbulence.
This can be a substantial fraction of the initial bulk flow energy
and such a field would be responding to the flow, not
the reverse.


To summarize: the physics of the launch region ($< 1AU$)  involves:
(1) Origin of large scale magnetic fields, field buoyancy to coronae, 
field relaxation into larger coronal structures,
(2) physics of centrifugal + magnetic 
acceleration of material from small to super-Alfv\'enic speeds,   
or Poynting flux driven bursts of acceleration,
(3) criteria for steady or  bursty jets, and 
(4) assessment of the extent of  Poynting flux domination.

The physics of the propagation region ($>> 1$AU observationally resolved)
involves such issues as:
(1) Propagation, instability formation, and sustenance of collimation in 
as a function of  internal vs. external density and  
strength of magnetic fields, 
(2) bow shocks, cocoon physics, particle acceleration, 
(3) effect of cooling on  morphology, and 
(4) interaction with ambient media



\section{Toward Connections between Theory and Observation}
\label{sec:5}

Spatially resolving the launch region and 
measuring the magnetic field strength and geometry 
therein would be the gold standard for directly evaluating the role of
magnetic fields in producing asymmetric p/PNe.
Measurements of fields in the propagation region 
 provide primarily indirect evidence, though the detection of relatively 
strong fields there is particularly  significant. \cite{vlemmings06}.

Whether binaries supply  needed
rotational energy to amplify jet-mediating fields in accretion disks is
a  fundamental question.
The basic wind kinematics are roughly consistent
with accretion onto a WD, suggesting the importance of accretion onto
the primary.  

Coupling accretion disk physics to large scale magnetic field production, to jet launch, and jet propagation in a unified theory 
is work in progress that spans several subfields of theoretical astrophysics,
 let alone the specific application to p/PNe.  However  some 
predictions/trends  can be studied:
(1) It should be possible to evaluate the kinematic constraints/predictions of Sec 3  in more detail and compare the distribution of inferred fast 
outflow speeds to what would be expected from 
known binary statistics of low mass main sequence stars. This would serve 
to help determine the commonality of  accretion onto the companion vs. the primary.
(2) CE models would  predict mostly  Oxygen rich rather than Carbon rich 
post AGB systems because typically, the companion ejects the envelope on time scales 
of order years, right from the beginning of the AGB phase when the envelope
expands, so the AGB star would not have had a chance to reach the
Carbon dredge up.
(3) Crystalline dust  in post-AGB systems can be  produced if a binary
induced spiral shock  anneals silicates \cite{edgar}. Is this universal?
(4) CE evolution would predict equatorial outflows
from companion inspiral that precedes any accretion driven
poloidal jet. A delay is observed \cite{huggins} but more work
is needed to predict the delay time scale.
(5) Are the geometry and composition of dust tori around  post-AGB objects
consistent with the influence of a binary?
(6) Are fast outflows contaminated by material that could represent 
accretion disk residue of  shredded low mass companions?
(7) Are time scales of observed outflow 
precession consistent with the gravitational
influence of a binary on a disk?
(8) Can  double peaked line profiles be detected
to identify accretion disks within the launch region?
(9) Can  shrouded novae outbursts from a $M_2>M_{crit}$ companion feeding the primary be detected in X-rays?
(10) Improved statistics on the fraction of
 bipolar pPNe, the fraction of suitable precursor binaries for CEE,
and the fraction of stars which evolve to be pPNe 
will constrain theories and 
assess whether all PNe incur asymmetry.

%
%
%
%
%
%

%
%




\begin{thebibliography}{99.}
%
%


\bibitem{bujar01} V. Bujarrabal, V., 
A. Castro-Carrizo, J. Alcolea, \& C. S{\'a}nchez Contreras,  \aap, 
{\bf 377}, 868 (2001)

\bibitem{vlemmings06} W.H.T. Vlemmings, P.J.Diamond,  \& H. Imai,
 Nature, 440, {\bf 58} (2006)

\bibitem{balickfrank02} B. Balick, \& A. Frank, ARAA, {\bf 40}, 439 (2002)

\bibitem{moe06} M. Moe \& O. De Marco, 
\apj, {\bf 650}, 916 (2006)


\bibitem{soker06} N. Soker,  \apjl, {\bf 645}, L57  (2006)



\bibitem{bfw01} E.G. Blackman, A.
Frank,\& C. Welch, \apj, {\bf 546}, 288  (2001)

\bibitem{ibenlivio} I.J. Iben,  \& M. Livio, 
PASP, {\bf 105}, 1373 (1993)


\bibitem{nb06} J. Nordhaus \& 
E.G. Blackman, \mnras, {\bf 370}, (2006)

\bibitem{rrl99} M. Reyes-Ruiz, \& J.A. L{\'o}pez, ApJ, {\bf 524}, 952 (1999) 

\bibitem{huggins} P.J. Huggins, \apj, 
{\bf 663}, 342 (2007)

\bibitem{gl06} D. Grether, \& 
C.H. Lineweaver, ApJ, {\bf 640}, 1051 (2006)

\bibitem{sl94} N. Soker, \& M. Livio, ApJ, {\bf 421}, 219 (2004)

\bibitem{mm98} N. Mastrodemos,
\& M. Morris\ 1998, ApJ, {\bf 497}, 303 (1998)


\bibitem{soker05} N. Soker, AJ, {\bf 129}, 947 (2005)


\bibitem{pudritz04} R.E. Pudritz, 
Les Houches Summer School, {\bf 78}, 187. (2004)

(2004)



\bibitem{blackman07} E.G. Blackman, Astrphys. Sp. Sci., 
{\bf 307}, 7, (2007)

\bibitem{matt06} S. Matt, A. Frank \& 
E.G. Blackman, \apjl, {\bf 647}, L45 (2006)

\bibitem{garcia05} 
G. Garc{\'{\i}}a-Segura, J.A.  L{\'o}pez, \& J. Franco, \apj, {\bf 618}, 
919 (2005) 


\bibitem{anderson} J.M. Anderson, Z.Y. Li, R.  Krasnopolsky,\& R.D. 
Blandford, 2006, \apjl, {\bf 653}, L33 (2006)


 






\bibitem{garcia97} G. Garcia-Segura, 
\apjl, {\bf 489}, L189 (1997)
 



\bibitem{dennis07} T.J. Dennis  A.J.Cunningham, A. Frank, B. Balick,
E.G. Blackman, \& S. Mitran  arXiv:0707.1641  submitted to ApJ (2007) 




\bibitem{nbf07}J. Nordhaus, E.G. 
Blackman, \& A. Frank, \mnras, {\bf 376}, 599 (2007)

\bibitem{blackman07b} E.G. Blackman, arXiv:0707.3191,
New. J. Phys, in press (2007)


\bibitem{edgar} R.G. Edgar, E.G. Blackman, J.T. Nordhaus, A. Frank,
in preparation (2007)






























\end{thebibliography}
\end{document}